\documentclass[a4paper]{jpconf}
\usepackage{graphicx}
\usepackage{bm}
\usepackage{amsmath}
\usepackage{amssymb}
\newcommand{\cprx}{CePd$_{1-x}$Rh$_{x}$}
\newcommand{\cncx}{CeNi$_{1-x}$Cu$_{x}$}
\newcommand{\cpr}{CePd$_{0.2}$Rh$_{0.8}$}

\begin{document}
\title{Quantum Griffiths phase in \cprx~with $x \approx 0.8$}
\author{M Brando$^{1}$, T Westerkamp$^{1}$, M Deppe$^{1}$, P Gegenwart$^{2}$, C Geibel$^{1}$ and F Steglich$^{1}$}
\address{$^1$ Max Planck Institute for Chemical Physics of Solids, Dresden 01178, Germany}
\address{$^2$ I. Physikalisches Institut, Georg-August-Universit\"at, G\"ottingen 37077, Germany}

\ead{manuel.brando@cpfs.mpg.de}
\begin{abstract}
The magnetic field dependence of the magnetisation ($M$) and the temperature dependence of the ac susceptibility ($\chi' = dM/dH$) of \cprx~single crystals with $0.80 \leq x \leq 0.86$ are analysed within the frame of the quantum Griffiths phase scenario, which predicts $M \propto H^{\lambda}$ and $\chi' \propto T^{\lambda-1}$ with $0 \leq \lambda \leq 1$. All $M$ vs $H$ and $\chi'$ vs $T$ data follow the predicted power-law behaviour. The parameter $\lambda$, extracted from $\chi'(T)$, is very sensitive to the Rh content $x$ and varies systematically with $x$ from -0.1 to 0.4. The value of $\lambda$, derived from $M(H)$ measurements on a~\cpr~single crystal, seems to be rather constant, $\lambda \approx 0.2$, in a broad range of temperatures between 0.05 and 2~K and fields up to about 10 T. All observed signatures and the $\lambda$ values are thus compatible with the quantum Griffiths scenario.
\end{abstract}
\section{Introduction}
Ce- and Yb-based $f$-electron Kondo-lattice (KL) systems have shown the most drastic forms of non-Fermi-liquid (NFL) behaviour~\cite{Loehneysen2007}. In these heavy-fermion (HF) materials, the ground state sensitively depends on the balance between two competing interactions, which are both determined by the strength of the $4f$-conduction electron hybridisation $J$: Whereas the Kondo interaction leads to a screening of the local moments below a Kondo temperature $T_{K}$, resulting in a paramagnetic (PM) ground state with itinerant $4f$-electrons, the indirect exchange coupling (RKKY interaction) can mediate long-range magnetic ordering~\cite{doniach1977}.\\
One of the explanations for such NFL phenomena is the presence of a quantum critical point (QCP) at a particular $J_{c}$: If the transition temperature $T_{M}$ of the long-range magnetic order is continuously shifted to zero by an external parameter $x(J)$, e.g. pressure, magnetic field or chemical substitution, a $2^{nd}$ order quantum phase transition (QPT) takes place at $x_{c}(J_{c})$ and $T = 0$, to which a QCP is associated. Here, the typical length and time scales of order parameter fluctuations diverge when approaching the transition point. These fluctuations are believed to be responsible for the observed NFL corrections to the FL prediction for the heat capacity $C/T(T) = const.$, magnetic susceptibility $\chi(T) = const.$, electrical resistivity $\rho(T) \propto T^{2}$ etc. QCPs are not the only mechanism to provide NFL behaviour. If the magnetic order changes from long-range  to short-range and disorder comes into play, spatial regions (also called ``rare regions'') can show local magnetic order, although the bulk system is in a PM state~\cite{vojta2006}. The order parameter fluctuations of these regions can become strong enough to destroy the QPT and can give rise to NFL effects~\cite{miranda1997,castroneto1998,castroneto2000}. Here, the length scale of order parameter fluctuations becomes finite, while the time scale still diverges. As a consequence in, e.g., the quantum Griffiths phase (QGP) scenario, power-law corrections, $C/T \propto \chi' \propto T^{\lambda - 1}$ and $M \propto H^{\lambda}$ with $0 \leq \lambda \leq 1$, are expected in a broad region across the QCP and not only at the QCP itself~\cite{vojta2005}. The global phase transition is then smeared~\cite{hoyos2008}, as observed in doped ferromagnetic (FM) materials, as , e.g., the itinerant Zr$_{1-x}$Nb$_{x}$Zn$_{2}$~\cite{sokolov2006}, the $5f$-based HF system URh$_{1-x}$Ru$_{x}$Ge~\cite{huy2007} and the Ce-based~\cprx~\cite{sereni2007}. The ground state of such systems depends on several factors and can be very exotic, as in~\cncx, where a percolative cluster scenario has been proposed~\cite{marcano2007}.\\
Recently, we reported on the formation of a ``Kondo-cluster-glass'' state in \cprx~for $0.7 \leq x \leq 0.9$, caused by the freezing of clusters with predominantly FM coupling: they form below a temperature $T_{cl}$ and freeze at a lower temperature $T_{c}^{*}$~\cite{westerkamp2009}. In~\cprx, the chemical substitution of the Ce-ligand Pd by Rh induces not just a negative volume effect, but, more importantly, it locally increases the hybridisation strength $J$ of the cerium $4f$ electrons, leading to a strong enhancement of the local $T_{K}$. Simultaneously, disorder is introduced to the system, which induces a statistical distribution of $T_{K}$. Both effects create regions where $f$ moments are still unscreened and can form FM clusters because of the RKKY interaction~\cite{dobrosavljevi2005}.
\begin{figure}[b]
\begin{center}
\includegraphics[width=15pc,angle=-90]{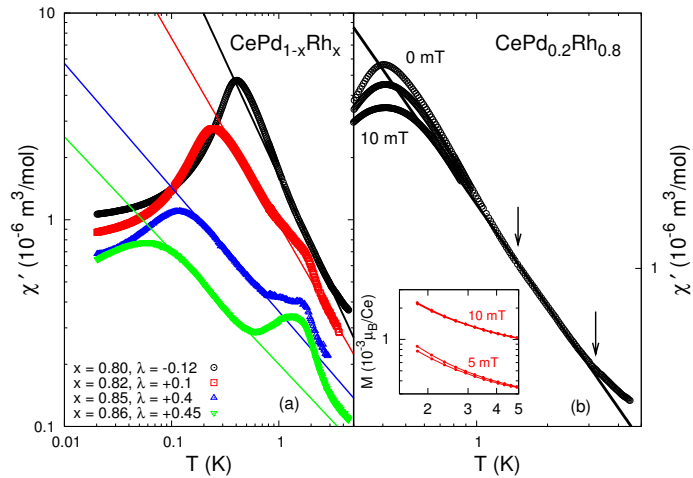}
\caption{\label{fig1} Frame (a): Ac susceptibility $\chi'$ of four single crystals plotted as a function of the temperature on a double-logarithmic scale. The lines indicate fits with $\chi' \propto T^{\lambda - 1}$ above $T_{c}^{*}$. The curves present a hump at about 3~K, which is possibly due to impurity phases. Frame (b): $\chi'$ vs $T$ of \cpr~at different magnetic fields: 0, 5 and 10~mT. The line represents a linear fit performed between 3.5~K and 1.5~K. Inset of frame (b): FC and ZFC measurements of the magnetisation at constant fields of 5 and 10mT down to 2~K. The FC and ZFC data splits slightly below $T_{cl}$ in 5~mT, where clusters form; a field of 10~mT is enough to remove this effect.}
\end{center}
\end{figure}
In the temperature region $T_{c}^{*} < T < T_{cl}$ we have observed power-law behaviour of several thermodynamic quantities, i.e. $C/T(T)$, $\chi'(T)$, $M(H)$~\cite{westerkamp2009,pikul2006}, and we concluded that the quantum Griffiths phase scenario might be realised in \cprx~for Rh content $0.8 \leq x \leq 0.9$. For $x < 0.8$ the system is too close to the FM long-range order to be considered in the QGP. In this article we take a closer look at the magnetic susceptibility $\chi'$ and magnetisation $M$ in single crystals with Rh content $x_{c} \approx 0.8$ within the frame of the QGP scenario.\\
Before starting with the analysis of the results, we have to consider that the QGP is restricted to a small region of the temperature - magnetic field ($T - H$) phase diagram. This region is limited to the ranges $T_{c}^{*} < T < T_{cl}$ and $H(T_{c}^{*}) < H < H(T_{cl})$, where $H(T_{c}^{*})$ and $H(T_{cl})$ are, respectively, the magnetic fields necessary to destroy the glass state and to remove the effect of the cluster formation (cf. inset of Fig.~\ref{fig1} (b)). Both $T_{c}^{*}$ and $H(T_{c}^{*})$ have very low values of the order of mK and mT. To study the dynamic processes in the region close to $x = 0.8$, $\chi'$ was measured down to 0.02~K (Fig.~\ref{fig1}), at a frequency of 113~Hz, with a modulation field of $\mu_{0}h_{ac} = 11~\mu$T. The dc magnetisation $M$ was measured with a SQUID (Quantum Design) at high temperatures and with a high-resolution Faraday magnetometer, in magnetic fields as high as 11~T and at temperatures down to 0.05~K~(Fig.~\ref{fig2} and~\ref{fig3})\cite{sakakibara1994}. The magnetic field has been applied along the $c$ axis in all the measurements presented here. It is worth mentioning that the magnetic anisotropy of the system is low for $x > 0.7$ ~\cite{deppe2006} and poly- and single crystals exhibit similar behaviour~\cite{westerkamp2009}.\\
In frame (a) of Fig.~\ref{fig1}, the real part of the ac susceptibility is plotted as a function of $T$ in a double-logarithmic scale for four crystals of~\cprx~with $0.8 \leq x \leq 0.86$. The peak temperature is $T_{c}^{*}$. 
\begin{figure}[b]
\begin{center}
\includegraphics[width=16pc,angle=-90]{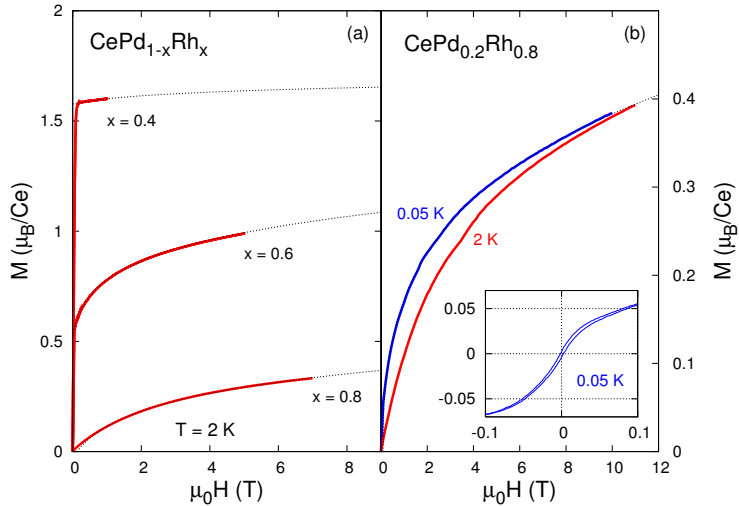}
\caption{\label{fig2} Frame (a): $M$ vs $H$ for three selected single crystals of \cprx~with $x$ = 0.4, 0.6 and 0.8 at 2~K. Dotted lines are guide to the eye. Frame (b): $M$ vs $H$ of \cpr~at 0.05 and 2~K. Inset of frame (b): By zooming into the field range $-0.1 \leq \mu_{0}H \leq 0.1$~T at 0.05~K, a very small FM hysteresis can be observed. The magnetic field was applied along the $c$ axis.}
\end{center}
\end{figure}
In frame (b), $\chi'$ is shown only for the crystal with $x = 0.8$ at different fields. $T_{cl} \approx 5$~K has been determined as the temperature where the field-cooled (FC) and zero-field-cooled (ZFC) measurements of $M$ split, as illustrated in the inset of the same frame. As in polycrystals, $\chi'$ follows a $T^{\lambda - 1}$ power-law function above $T_{c}^{*}$ where the exponent $\lambda$ varies systematically with $x$ between -0.12 and 0.45~\cite{westerkamp2009}, as expected in the QGP scenario. The value of the exponents seems to be independent of the impurity phases observed at about 3~K (see curve for $x = 0.82$). The negative $\lambda$ value for the sample with $x = 0.8$, different from the one measured in a polycrystal with the same $x$ ($\lambda = 0.13$), suggests that this sample is still too close to the FM instability. Moreover, the error in estimating the composition by energy dispersive X-ray spectrometry amounts to 1 at. \% Rh. Taking into account that the sample with slightly higher Rh content $x = 0.82$ shows $\lambda = 0.1$, it is not possible from this measurement to discern whether the sample with $x = 0.8$ can be considered in the QGP or not. Furthermore, the exponent is affected by the freezing which already takes place at $T_{c}^{*} \approx 0.4$~K. Focussing on the temperature range between $T_{c}^{*}$ and $T_{cl}$, the $\chi'$ vs $T$ plot changes slope slightly at about 3.5~K, and again at about 1.5~K (indicated by arrows). We have fitted this range to extract $\lambda$, considering that in the 10~mT curve the fit range can be expanded to even lower temperatures.\\
To verify the presence of a QGP in this sample, we have measured $M$ vs $H$ at different temperatures. In Fig.~\ref{fig2} the isotherms are shown: In frame (a), the curve at 2~K is plotted together with those for two FM concentrations, $x = 0.4$ and $0.6$, to emphasise the strong decreasing of the magnetic moment with $x$; in frame (b), the isotherms at 0.05~K and 2~K are compared. The magnetisation at 0.05~K shows a very small hysteresis (inset of Fig.~\ref{fig2}), due to freezing at $T_{c}^{*} \approx 0.4$~K, and it is far from reaching saturation at 10~T. We consider the value of the coercive field to be $H(T_{c}^{*}) \approx 0.1$~T. Since $T_{cl} \approx 5$~K~$\approx \mu\cdot H$, we can assume $H(T_{cl})$ to be close to 12~T, as the magnetic moment $\mu$ at 12~T is only $0.4~\mu_{B}$. It is thus plausible to look for QGP in $M$ vs $H$ for fields between 0.1 and 12~T. $M$ vs $H^{\lambda}$ is plotted in Fig.~\ref{fig3} at three temperatures; below $T_{c}^{*}$ (0.05~K), just above it (0.5~K) and close to $T_{cl}$ (2~K).
\begin{figure}[t]
\includegraphics[width=13pc,angle=-90]{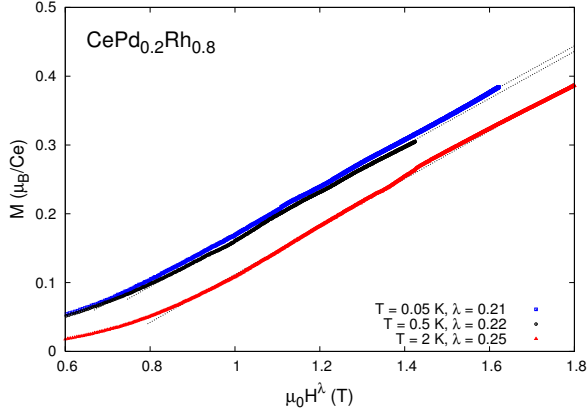}
\begin{minipage}[t]{18pc}\caption{\label{fig3} Magnetisation of \cpr~plotted as a function of $H^{\lambda}$ at three different temperatures: 0.05, 0.5 and 2 K. The linear behaviour of the three curves for fields $0.4 \lesssim \mu_{0}H \lesssim 10$~T indicate that the power-law $M \sim H^{\lambda}$ holds with an almost constant $\lambda$ in this entire field range. Linear dotted lines are just guide to the eye.}
\end{minipage}

\end{figure}
The curves follow a power-law behaviour, with an almost constant value of $\lambda \approx 0.2$, close to those observed in $\chi'(T)$ for the other crystals, as expected in the QGP. Since $M$ vs $H$ plots are less sensitive to paramagnetic impurities, and for $H > H(T_{c}^{*})$ the freezing does not affect the power law, these plots can be considered as signatures of a QGP in~\cpr.\\
As discussed in the introduction, there is a fundamental difference between the NFL behaviour given by long-range and short-range order fluctuations. The presence of clusters and the power-law corrections to susceptibility and magnetisation in~\cprx~indicate that the expected QPT at $x_{c} \approx 0.8$ is replaced by disordered phases, possibly like the Griffiths one.\\
We wish to thank T. Vojta for helpful discussions. Supported by the DFG Research Unit 960 "Quantum phase transition".

\end{document}